\documentstyle[preprint,eqsecnum,aps]{revtex}

\newcommand{\bphi}{\mbox{\boldmath $\phi$}}
\newcommand{\bdx}{\mbox{\boldmath $x$}}

\begin{document}
\draft
\preprint{hep-th/9711129}
\title{Variational calculation of the effective action}
\author{Takanori Sugihara
\footnote{e-mail : taka@rcnp.osaka-u.ac.jp}}
\address{Research Center for Nuclear Physics, 
Osaka University, Osaka 567, Japan}
\date{November 17, 1997}
\maketitle
\begin{abstract}
An indication of spontaneous symmetry breaking is found 
in the two-dimensional $\lambda\phi^4$ model, 
where attention is paid to the functional form of 
an effective action. 
An effective energy, 
which is an effective action for a static field, 
is obtained as a functional of the classical field
from the ground state of the hamiltonian $H[J]$ 
interacting with a constant external field. 
The energy and wavefunction of the ground state are 
calculated in terms of DLCQ 
(Discretized Light-Cone Quantization) 
under antiperiodic boundary conditions. 
A field configuration 
that is physically meaningful is found as a solution 
of the quantum mechanical Euler-Lagrange equation 
in the $J\to 0$ limit. 
It is shown that there exists a nonzero field configuration 
in the broken phase of $Z_2$ symmetry 
because of a boundary effect. 
\end{abstract}

\vspace{0.5cm}

\vfill\eject

\section{Introduction}
In order to explain properties of hadrons, 
it is strongly hoped that QCD particle spectra can be calculated 
with a reasonable approximation. 
Light-front field theory is one of the candidates 
to investigate QCD in the infrared region, 
since a certain nonperturbative approximation 
(Tamm-Dancoff truncation)
becomes effective \cite{phw,tamm_dancoff,mp,gauge2}. 
Since this method is based on a hamiltonian formalism, 
one can obtain mass spectra and wavefunctions 
of hadronic states 
which are important for the calculation of 
nonperturbative physical quantities 
such as structure functions \cite{bpp}. 

It has been said that 
spontaneous chiral symmetry breaking is responsible for 
the finite masses of mesons in the chiral limit. 
We need to know how the chiral condensate 
$\langle \bar{\psi}\psi \rangle$ behaves 
in the chiral region to understand QCD mesons. 
However, 
one cannot extract information 
of spontaneous symmetry breaking from the vacuum, 
since the light-front vacuum is always trivial. 
So the question is: 
How can we understand spontaneous symmetry breaking 
in the light-front field theory? 
It is usual to apply the method of the effective potential 
to such a problem. 
In order to define the effective potential on the light-front, 
let us consider a Legendre transform 
(which we call the effective energy) 
of the ground state energy of the following hamiltonian
\cite{kugo,sc}, 
\begin{equation}
  H[J]=H-\int d^{n-1}x J({\bdx})\phi({\bdx}), 
  \label{hj}
\end{equation}
where the external field is independent of time and 
{\bdx} indicates the spatial coordinate. 
The advantage of this consideration is that 
it is possible to obtain the effective energy 
if we could know only the ground state of the hamiltonian $H[J]$. 
If the external field does not depend on 
the spatial coordinate, 
the effective potential is given as the effective energy 
divided by the total spatial volume of the system. 
Then, it would be natural to define the system 
in a finite box $-L<{\bdx}<L$ 
and take the thermodynamic limit $L\to\infty$ 
after all calculations. 
This is known as 
DLCQ (Discretized Light-Cone Quantization) \cite{my,pb}. 
There are two possibilities 
for the consistent boundary condition on the field $\phi({\bdx})$:
periodic and antiperiodic boundary conditions \cite{ty}. 
If we take a periodic boundary condition and 
assume a uniform external field $J({\bdx})=J$, 
it would be possible in principle 
to obtain the effective potential from the effective energy. 
To do that, 
we have to know the light-front longitudinal zero mode 
\cite{my,kty,ty}, 
\begin{equation}
  \phi_0 = \frac{1}{2L}\int_{-L}^L dx^- \phi({\bdx}), 
\end{equation}
which appears in the second term of (\ref{hj}). 
If we impose a periodic boundary condition on the field, 
a constraint equation for the zero mode emerges. 
The light-front zero mode is a dependent variable and then 
should be represented with other oscillator modes. 
It has been numerically confirmed, with an approximation, 
that the zero mode gives rise to a non-zero 
vacuum expectation value and the entire effect of 
spontaneous symmetry breaking comes from only one mode \cite{zero}. 
It seems that this scenario holds also in the calculation 
of the light-front effective potential. 
The zero mode should have a singular dependence 
on the external field $J$ to produce a correct 
convex shape for the effective potential in the broken phase.
However, it does not work in practice 
solving the constraint and 
calculating the vacuum energy including the zero-mode effect, 
since the constraint equation is highly complicated and 
it is difficult to find some reasonable technique 
to solve it accurately. 
It is worthwhile to discuss the problem 
without the zero mode. 

The vacuum expectation value which minimizes 
the effective potential is a particular solution 
of the following 
quantum mechanical Euler-Lagrange equation 
\begin{equation}
  \frac{\delta\Gamma[\varphi]}{\delta\varphi(x)}=J(x), 
  \quad 
  J(x)\to 0, 
  \label{el0}
\end{equation}
where $\Gamma[\varphi]$ is an effective action. 
Of course, there should exist also a space-dependent solution 
and it would be possible to see indications of 
symmetry breaking in it. 
Our aim is to extract convexity of the effective potential 
from the hamiltonian searching for a non-uniform solution 
of Eq. (\ref{el0}). 
This will be done assuming an antiperiodic boundary condition 
and using a trick on the external field. 
Since the translational invariance is broken by the external field, 
we can deal with the problem by avoiding the zero mode. 
The translational invariance of the system is restored 
by taking the $J\to 0$ limit after all the calculations. 

In this paper, we look for an indication of 
spontaneous symmetry breaking 
in the two-dimensional $\lambda\phi^4$ model 
by paying attention 
to a functional form of the effective energy, 
where the system is defined and solved using DLCQ \cite{my,pb}, 
which is essentially a nonperturbative method and 
useful also as a conceptual tool \cite{ls}. 
The effective energy is obtained as a functional of 
the classical field (the expectation value of the field operator) 
and a space-dependent non-uniform solution of 
the quantum mechanical Euler-Lagrange equation is found. 

The essential point of this consideration is 
the imposition of an external field on the system. 
In order to describe the broken phase properly, 
we have to break the symmetry explicitly 
by imposing an external field on the system \cite{kty,ty}. 
The vanishing limit of the external field has to be taken 
after the thermodynamic limit. 
It is impossible to figure out the properties of 
the effective energy if the order of the limits is changed. 
In Ref. \cite{conv1,conv2}, it has been shown 
that the second derivative of the effective potential 
is always positive and in particular the potential does not exist 
for small expectation values 
if spontaneous symmetry breaking occurs. 
If the symmetry breaks, 
the effective potential should have a flat bottom 
and the finite expectation value of the field 
survives in the $J\to 0$ limit. 
The effective energy would also have a convex shape 
as a function of the classical field $\varphi({\bdx})$, 
since the energy is a more general quantity than the potential 
and should contain information of the potential. 

This paper is organized as follows. 
In Sec. \ref{sec_action}, 
the effective energy is defined in terms of a hamiltonian 
that interacts with an external field $J({\bdx})$. 
It is explained how to obtain a physically meaningful 
field configuration $\varphi({\bdx})$. 
In Sec. \ref{sec_dlcq}, 
the DLCQ method is introduced under an antiperiodic condition  
to solve the eigenvalue problem given by the hamiltonian. 
An approximate value of the critical coupling constant 
is calculated in a nonperturbative manner. 
Mass spectra for periodic and antiperiodic boundary conditions 
are compared with each other for reference. 
In Sec. \ref{sec_energy} 
the effective energy is calculated 
using DLCQ introduced in Sec. \ref{sec_dlcq}. 
We will see that 
there seems to remain a nonzero configuration 
in the broken phase as a solution 
of the Euler-Lagrange equation 
even if the external field is switched off. 
Sec. \ref{sec_sum} is devoted to summary and discussions. 

\section{Effective action and hamiltonian}
\label{sec_action}
Let us consider a generating functional $Z[J]$ of the Green's function 
to define the effective energy in terms of the hamiltonian 
\cite{kugo,sc}, 
\begin{equation}
  Z[J]=e^{iW[J]} =
  \langle 0 | T \exp\left[\int d^n x J(x)\phi(x)\right] | 0 \rangle. 
  \label{z}
\end{equation}
If the external field $J(x)$ is independent of time, 
the partition function can be written with a hamiltonian $H[J]$ 
that interacts with the external field in the following way, 
\begin{equation}
  Z[J] = e^{-iw[J]T} = \langle 0 | e^{-iH[J]T} | 0 \rangle, 
  \label{z2}
\end{equation}
\begin{equation}
  H[J] = H - \int d^{n-1}x J({\bdx})\phi({\bdx}), 
  \label{hj2}
\end{equation}
where ${\bdx}$ means spatial coordinate 
and $\phi({\bdx})$ is a field operator 
in the Schr\"odinger picture. 
The proof of (\ref{z2}) is given in Appendix \ref{p_h}. 
In the relation (\ref{z2}), it is understood that 
the $-i\epsilon$ prescription is taken, that is, 
the time coordinate is rotated with a replacement 
$H[J]\to e^{-i\epsilon}H[J]$ ($\epsilon \ll 1$). 
By substituting the decomposition of unity into (\ref{z2}) 
and taking the $T\to\infty$ limit, 
we can see that the ground state $|0_J\rangle$ 
of the hamiltonian $H[J]$ dominates in $Z[J]$, 
\begin{equation}
  H[J]|0_J\rangle = w[J]|0_J\rangle, 
  \label{esj}
\end{equation}
where the state is normalized as $\langle 0_J|0_J \rangle=1$. 
The connected generating functional $w[J]$ can be regarded as 
the ground state energy of the hamiltonian $H[J]$. 
By multiplying Eq. (\ref{esj}) by the ground state $\langle 0_J|$, 
we have 
\begin{equation}
  \langle 0_J | H | 0_J \rangle 
  = w[J] + \int d^{n-1}x J({\bdx})\varphi({\bdx}), 
  \label{eff}
\end{equation}
where 
\begin{equation}
  \varphi({\bdx}) = \langle 0_J | \phi({\bdx}) | 0_J \rangle. 
\end{equation}
Since (\ref{eff}) is a Legendre transform of $w[J]$, 
this quantity is just an effective action divided by the 
total time $T$ in the case when the field $\varphi(x)$ is static, 
\begin{equation}
  \Gamma[\varphi(x)=\varphi({\bdx})]=
  -T{\cal E}[\varphi({\bdx})]. 
\end{equation}
We call this quantity (\ref{eff}) the effective energy 
${\cal E[\varphi]}$, 
\begin{equation}
  {\cal E}[\varphi] \equiv \langle 0_J | H | 0_J \rangle. 
\end{equation}
An actual expectation value $\varphi({\bdx})$ 
of the field operator $\phi({\bdx})$ should be given as a solution 
of the following generalized Euler-Lagrange equation 
with vanishing external field, 
\begin{equation}
  \frac{\delta {\cal E}[\varphi]}{\delta \varphi({\bdx})} 
  = J({\bdx}), 
  \quad
  J({\bdx})\to 0. 
  \label{el}
\end{equation}
In order to obtain the solution $\varphi({\bdx})$, 
we have to make three steps:
(1) solve the eigenvalue problem (\ref{esj}), 
(2) evaluate the energy ${\cal E[\varphi]}$, 
(3) find the stationary point of ${\cal E[\varphi]}$. 
It is difficult to clear the first step 
if the field is quantized 
in the ordinary equal-time coordinate, 
because vacuum fluctuations dominate and 
a higher Fock state seems to be needed to 
represent the ground state of the hamiltonian. 
In order to solve Eq. (\ref{esj}) 
in a nonperturbative manner with a reasonable approximation, 
DLCQ (Discretized Light-Cone Quantization) 
will be used in Sec. \ref{sec_energy}. 
The effective energy ${\cal E}[\varphi]$ will be obtained 
as a functional of the classical field $\varphi(\bdx)$ 
by diagonalizing the hamiltonian $H[J]$. 

\section{Critical coupling constant with DLCQ}
\label{sec_dlcq}
\subsection{DLCQ in the $\lambda\phi_{1+1}^4$ model}
In this section, we will consider DLCQ \cite{my,pb} 
in order to apply the method introduced in the previous section 
to the two-dimensional real scalar model \cite{chang,hv,xu,hswz}. 
An approximate value of the critical coupling constant 
$\lambda_{\rm c}$ will be calculated. 

The Lagrangian density of the model is given by 
\begin{equation}
  {\cal L} = \frac{1}{2}
  \left( \partial_\mu\phi \partial^\mu\phi - \mu^2 \phi^2 \right)
  - \frac{\lambda}{4!}\phi^4. 
\end{equation}
The light-front coordinate 
$x^\pm = (x^0\pm x^1)/\sqrt{2}$
is defined and $x^+$ and $x^-$ are 
regarded as time and space, respectively. 
The metric is $g^{+-}=g^{-+}=1$ and  
$g^{++}=g^{--}=0$. 
The system is put in a finite size box ($-L \le x^- < L$) 
and the field is quantized with 
an antiperiodic boundary condition $\phi(L)=-\phi(-L)$ \cite{ty}, 
\begin{equation}
  [\phi(x),\phi(y)]_{x^+=y^+} = -\frac{i}{4}\epsilon(x^- - y^-). 
\end{equation}
The field is expanded with oscillators at $x^+=0$, 
\begin{equation}
  \phi(x)|_{x^+=0} = \frac{1}{\sqrt{4\pi}}
  \sum_{n=1}^\infty \frac{1}{\sqrt{\tilde{n}}} 
  \left[
    a_n e^{-ik^+_n x^-} + a_n^\dagger e^{ik^+_n x^-}
  \right], 
\end{equation}
where 
\begin{equation}
  k_n^+=\frac{\pi\tilde{n}}{L}, \quad
  \tilde{n}=n-\frac{1}{2}, 
\end{equation}
and 
\begin{equation}
  [a_m,a_n^\dagger]=\delta_{m,n}, \quad 
  [a_m,a_n]=0, \quad [a_m^\dagger,a_n^\dagger]=0. 
\end{equation}
The hamiltonian and momentum operators are 
\begin{equation}
  H = \int_{-L}^L dx^- :T^{+-}:, 
  \quad
  P = \int _{-L}^L dx^- :T^{++}:, 
\end{equation}
where 
\begin{equation}
  T^{\mu\nu}=\partial^\mu \phi \partial^\nu \phi - g^{\mu\nu} {\cal L}. 
\end{equation}
Divergent tadpole diagrams are removed by normal-ordering. 
The size of the box $L$ can be extracted from $H$ and $P$ 
\begin{equation}
  H = \frac{L}{2\pi}{\cal H}, 
  \quad
  P = \frac{\pi}{L}{\cal K}. 
\end{equation}
Explicit forms of ${\cal H}$ and ${\cal K}$ are 
written in appendix \ref{app_a}. 
The invariant mass $M$ of a state is 
\begin{equation}
  M^2 = {\cal KH}. 
  \label{mass2}
\end{equation}
Note that the invariant mass does not depend on $L$. 
The harmonic resolution ${\cal K}$ has to be taken to infinity 
after all calculations so as to give a finite fixed momentum $P$ 
in the thermodynamic limit $L\to\infty$. 
Then, we can say that $M$ depends on $L$ implicitly. 

Since $H$ and $P$ commute with each other, 
we can diagonalize these operators simultaneously. 
It is convenient to expand a general state 
as an eigenstate of ${\cal K}$
\begin{equation}
  {\cal K}|K\rangle = K|K\rangle. 
\end{equation}
where 
\begin{equation}
  |K\rangle = 
  \lim_{N_{\rm TD}\to\infty}
  \sum_{N=1}^{N_{\rm TD}} \sum_{n_1,n_2,\dots,n_N}^K
  \delta_{\sum_{i=1}^N n_i,K} c_{n_1,n_2,\dots,n_N}
  |n_1,n_2,\dots,n_N\rangle, 
  \label{kstate}
\end{equation}
and
\begin{equation}
  |n_1,n_2,\dots,n_N \rangle \equiv 
  \prod_{i=1}^N a_{n_i}^\dagger |0 \rangle. 
\end{equation}
The Fock space is truncated by 
the number of particles $N_{\rm TD}$ and 
the harmonic resolution $K$ 
in actual calculations, 
since the number of states ${\cal N}_{\rm s}$ goes to infinity 
in the limits $N_{\rm TD}\to\infty$ and $K\to\infty$ 
and it is impossible to manipulate 
infinite dimensional matrices. 
Invariant mass $M$ and wavefunction $c_i$ are obtained 
by diagonalizing the finite dimensional matrix 
${\cal H}_{ij}$ \cite{lapack} 
\begin{equation}
  M^2 c_i = K \sum_{j=1}^{{\cal N}_{\rm s}} {\cal H}_{ij}c_j, 
\end{equation}
where 
\begin{equation}
  {\cal H}_{ij}=\langle i|{\cal H}|j \rangle, 
  \quad
  |K\rangle = \sum_{i=1}^{{\cal N}_{\rm s}} c_i|i\rangle. 
  \label{state}
\end{equation}
After this, all quantities which have mass dimension 
will be expressed in units of $\mu^2$ 
due to the absence of transverse component in this model. 

The harmonic resolution $K$ is 
a total sum of $N$-particle momenta, 
each of which carries a half-integer piece, 
$\tilde{n}_i=n_i-1/2$, 
\begin{equation}
  K = \sum_{i=1}^N \tilde{n}_i, 
\end{equation}
then we have 
\begin{equation}
  \sum_{i=1}^N n_i = K + \frac{N}{2}. 
  \label{sum}
\end{equation}
Since the left-hand side of (\ref{sum}) is always integer, 
the number of particles $N$ should be odd or even 
according to whether $K$ is half-integer or integer, 
respectively. 
Then, odd and even sectors decouple from each other. 
The resolution $K$ is set to be half-integer, 
because our purpose in this section is to obtain 
a mass spectrum of the lightest particle state, 
which belongs to the odd sector and 
can be seen as a one-bosonic state. 

\subsection{Critical coupling constant}
The Tamm-Dancoff dependence of the lightest mass 
is shown in Fig. \ref {td}. 
The mass squared $M^2$ of the lightest state 
is plotted as a function of harmonic resolution $K$ 
in a case $\lambda=25\mu^2$, which is comparatively large 
and near to the critical coupling constant $\lambda_{\rm c}$. 
This state can be seen as a one-bosonic state, 
since one body component of wavefunction is dominant. 
A definite value of $\lambda_{\rm c}$ will be calculated later. 
Diamonds, pluses, squares, and crosses correspond to 
$N_{\rm TD}=3,5,7,9$, respectively, 
where $N_{\rm TD}$ is defined in (\ref{kstate}). 
Since the coupling constant is large, 
convergence with respect to $K$ and $N_{\rm TD}$ is very slow. 
The spectrum seems almost to converge at $N_{\rm TD}=7$ 
in the small $K$ region. 
The harmonic resolution $K$ cannot be taken to be large values 
when $N_{\rm TD}$ is large 
because of upper bounds of computational resources, 
especially a shortage of memory size. 
We will continue our calculations with $N_{\rm TD}=5$ 
and obtain the critical coupling constant $\lambda_{\rm c}$ 
by extrapolating the results, 
because the purpose of this section is 
to prepare a nonperturbative technique for 
the effective energy calculation and to find an approximate value 
of the critical coupling constant $\lambda_{\rm c}$ 
that is needed to draw a phase diagram. 
The effective energy for both phases 
will be evaluated in the next section using DLCQ. 

The $K$ dependence of the one-bosonic masses is shown 
in Table \ref{kdep_anti} for various coupling constants, 
where the number of particles 
is truncated with $N_{\rm TD}=5$. 
Since the mass spectra for large coupling 
do not converge to the extent of this calculation, 
let us estimate where the spectra settle in the $K\to\infty$ limit 
by expanding the mass squared with $1/K$ up to the second order, 
\begin{equation}
  M^2(\lambda) =
  m_0 + \frac{m_1}{K} + \frac{m_2}{K^2}. 
  \label{ext}
\end{equation}
The coefficients $m_i(\lambda)$ are obtained 
with least squares fitting 
using Marquardt-Levenberg algorithm.  
The result is shown in Table \ref{ml}. 
$m_0$ is regarded as an extrapolated value 
in the $K\to\infty$ limit. 

In Fig. \ref{mass}, 
the lightest masses $M^2$ of two cases 
(antiperiodic and periodic boundary conditions) 
are compared with each other. 
Extrapolated values $m_0$ of the lightest mass squared 
are plotted as functions of the coupling constant. 
In the periodic case, 
the calculation has been executed tentatively 
excluding the zero mode from the hamiltonian. 
By fitting the points with curves and 
extrapolating the curves in Fig. \ref{mass}, 
the approximate value of the critical coupling constant 
$\lambda_{\rm c}$ is found as 
\begin{equation}
  \lambda_{\rm c}
  = 
  \cases{
    28.6329 \mu^2 & (antiperiodic) \cr
    30.8431 \mu^2 & (periodic) \cr
    }. 
  \label{crit}
\end{equation}
$\lambda_{\rm c}$ is defined as the point 
that gives a massless eigenvalue $M^2(\lambda_{\rm c})=0$. 
These values are nothing but upper bounds of the true value, 
because DLCQ is a variational method and the size of the 
variational space cannot be taken to infinity. 
The results (\ref{crit}) are consistent with the values 
$22\mu^2<\lambda_{\rm c}<55\mu^2$ obtained 
in the conventional equal-time theory \cite{chang}, 
but much smaller than 
$\lambda_{\rm c}=4\pi(3+\sqrt{3})\mu^2\sim 59.5\mu^2$ \cite{zero}, 
$\lambda_{\rm c}=43.9\mu^2$ \cite{hv}, and 
$\lambda_{\rm c}=40\mu^2$ \cite{hswz}. 
A convergence of the spectrum in the antiperiodic case is 
slightly faster than the periodic one. 
The spectra are similar 
but we cannot conclude clearly 
whether the two results coincide or not, 
since the spectra are extrapolated 
to the large $K$ region in this calculation. 
It is interesting to see how the zero-mode effect dominates 
in the mass spectrum calculation 
to confirm the equivalence between both boundary conditions. 
It seems that a certain renormalization technique 
needs to be found to get a convergent result 
with small $K$ \cite{sande}.

\section{Effective energy}
\label{sec_energy}
\subsection{Kink solution}
In this section, 
we will calculate the effective energy of 
the two-dimensional real scalar model with DLCQ 
and obtain an expectation value $\varphi(x^-)$ as 
a solution of (\ref{el}). 
It is possible to understand all the static physics of the system 
once a solution of the Euler-Lagrange equation 
(\ref{el}) is obtained. 
The configuration $\varphi(x^-)$ must contain 
information of spontaneous symmetry breaking 
independent of whether $\varphi(x^-)$ is uniform or not. 
If we impose a periodic boundary condition on the field 
$\phi(L)=\phi(-L)$ and assume a uniform external field $J(x^-)=J$, 
the effective potential ${\cal V}(\varphi)$ is obtained 
as the effective energy divided by the total spatial volume $2L$, 
\begin{eqnarray}
  {\cal V}(\varphi)
  &\equiv&
  \frac{1}{2L}{\cal E}[\varphi(x^-)=\varphi]
  \nonumber
  \\
  &=&
  \frac{1}{2L}w(j) + J \langle 0_J|\phi_0|0_J \rangle, 
\end{eqnarray}
where $\phi_0$ is a zero-mode part 
of the field operator $\phi(x^-)$ 
\begin{equation}
  \phi_0 \equiv \frac{1}{2L} \int_{-L}^L dx^- \phi(x^-). 
\end{equation}
A vacuum expectation value of the field operator 
$\varphi=\langle 0_J|\phi(x^-)|0_J\rangle$ is 
given as a solution of the following equation, 
\begin{equation}
  \frac{d{\cal V}(\varphi)}{d\varphi} = J, \quad J\to 0, 
\end{equation}
where $\varphi$ is independent of space 
because of translational invariance of the system. 
In order to evaluate the potential, 
we have to consider a constrained zero mode $\phi_0$, 
which is given as a solution of 
the following constraint equation \cite{my}, 
\begin{equation}
  \int_{-L}^L dx^-
  \left(
    \mu^2\phi + \frac{\lambda}{6}\phi^3 - J
  \right)
  =0. 
  \label{zero}
\end{equation}
It is expected that the zero mode $\phi_0$ has a singular behavior 
with respect to the external field $J$ in the broken phase 
due to the convexity of the effective potential 
\cite{conv1,conv2}. 
However, it is difficult to solve Eq. (\ref{zero}) 
and represent the zero mode as a superposition of other modes, 
since it is an operator valued nonlinear equation. 
It would be better if we could understand a mechanism of 
spontaneous symmetry breaking without such a complicated 
zero-mode problem. 

Let us look for a non-uniform solution 
to avoid the zero-mode problem. 
If we impose an antiperiodic boundary condition on the field 
and assume a constant external field $J(x^-)=J$, 
a non-uniform solution $\varphi(x^-)$ 
that has a kink will be obtained 
because the external field has a discontinuity 
at the boundary $x^-=\pm L$. 
There cannot exist a translationally invariant solution, 
since the system has been connected at the boundary with a twist. 
The purpose of this section is to see 
whether there exists a nonzero solution $\varphi(x^-)$ of 
(\ref{el}) in the limit $J\to 0$ after all of the calculations. 

Since the system interacting with the constant external field 
is not translationally invariant 
under antiperiodic boundary condition, 
we cannot diagonalize the hamiltonian and momentum operators 
at the same time $[H[J],P]\ne 0$. 
A general state of eq. (\ref{esj}) should be expanded 
as a superposition of various momentum states 
\begin{equation}
  |\Psi\rangle = c_0|0\rangle + 
  \lim_{K_{\rm cut}\to\infty}
  \sum_{n=1}^{2K_{\rm cut}} |K=n/2\rangle, 
  \label{state2}
\end{equation}
where the resolution takes both half-integer and integer 
$K=1/2,1,3/2,2,\dots,K_{\rm cut}$ 
and $K_{\rm cut}$ is set to be some finite value 
that leads to convergence in the spectra. 
The odd and even sectors interact with each other 
due to the existence of a translationally non-invariant 
interaction, which can be observed in (\ref{int_j}). 

It is possible to confirm 
that the state (\ref{state2}) is expressed 
with a complete set of momenta 
by taking the continuum limit $L\to \infty$. Then 
\begin{equation}
  \hat{P}|\Psi\rangle = \sum_{K=0}^\infty 
  \left( \frac{\pi}{L}K \right) |K\rangle, 
\end{equation}
becomes 
\begin{equation}
  \hat{P}|\Psi\rangle = \int_0^\infty dP P |P\rangle, 
\end{equation}
where 
\begin{equation}
  |P\rangle \equiv
  \sum_{N=0}^\infty
  \int\left[\prod_{i=1}^N dp_i \right]
  \delta\left(\sum_{i=1}^N p_i - P\right)
  \psi_N(p_1,p_2,\dots,p_N)
  \left[\prod_{i=1}^N a^\dagger (p_i)\right]
  |0\rangle, 
\end{equation}
and
\begin{equation}
  \psi_N(p_1,p_2,\dots,p_N) \equiv
  \frac{L^{N/2}}{2^{(N+2)/2}\pi^{N+1}}
  c_{n_1,n_2,\dots,n_N}. 
\end{equation}

In order to calculate the effective energy, 
an eigenvalue problem 
\begin{equation}
  H[J]|\Psi\rangle = w[J]|\Psi\rangle, 
\end{equation}
is solved by numerical diagonalization 
of the hamiltonian ${\cal H}[J]=(2\pi/L)H[J]$. 
By substituting the energy $w[J]$ and wavefunction $|0_J\rangle$ 
of the ground state into the following relation
\begin{equation}
  \langle 0_J |{\cal H}[J]|0_J \rangle = \frac{2\pi}{L}w[J], 
\end{equation}
the effective energy is given as 
\begin{equation}
  \frac{2\pi}{L}{\cal E}[\varphi]
  = \frac{2\pi}{L}w[J] + 
  \sum_{n=1}^\infty 
  \langle 0_J| f_n(-L) a_n^\dagger + f_n^*(-L) a_n |0_J \rangle, 
  \label{eff2}
\end{equation}
where coefficients $f_n$ are defined in (\ref{f_n}). 
The spatial integration has been performed 
before evaluating contractions of the operators 
in the second term of the right hand side. 
We can obtain the left hand side as a functional of 
the classical field 
$\varphi(x^-)=\langle 0_J|\phi(x^-)|0_J \rangle$, 
which can be also calculated by using 
the wavefunction of the ground state $|0_J\rangle$.

\subsection{Numerical result}
\label{num}
In Fig. \ref{energy}, 
the generating functional $w[J]$ and 
the effective energy ${\cal E}[\varphi]$ 
are plotted for various $N_{\rm TD}$'s 
as functions of the harmonic resolution $K_{\rm cut}$ 
to check the convergence with respect to $K_{\rm cut}$, 
where $\lambda=50\mu^2$ and $J=0.01$ are taken. 
The effective energy has been obtained by evaluating 
the right hand side of (\ref{eff2}). 
That is, the Legendre-transform is numerically performed 
in terms of the eigenvalue $w[J]$ and 
the wavefunction $|0_J\rangle$ of the ground state of 
the hamiltonian $H[J]$. 
After this, 
we will take a parameter set $K_{\rm cut}=31/2$ 
and $N_{\rm TD}=7$ because this parameter set seems to 
give almost convergent spectra. 
The convergence of the spectra is slightly faster 
than mass spectrum calculation, 
because an expansion of a state $|\Psi\rangle$ 
starts from the zero-body state $|0\rangle$. 

In Fig. \ref{phi_x}, 
the classical field $\varphi(x^-)$ is plotted 
as a function of the spatial coordinate 
both in (a) symmetric ($\lambda=0.1\mu^2<\lambda_{\rm c}$) 
and (b) broken ($\lambda=50\mu^2>\lambda_{\rm c}$) phases, 
where the external field is changed 
at a regular interval of $\Delta J=0.05$. 
We can see that 
the field configuration $\varphi(x^-)$ is nearly uniform 
except at the boundary and has a twist at $x^-=\pm L$ 
because of the antiperiodicity of the field operator $\phi(x^-)$. 

In order to see how the magnitude of the classical field 
behaves with changing $J$, 
the $J$ dependence of the classical field $\varphi(x^-)$ 
is shown in Fig. \ref{jphi}. 
The maximum value $\varphi_{\rm max}\equiv{\rm max}\{\varphi(x^-)\}$ 
of the classical field is plotted as a function of $J$
both in symmetric ($\lambda=0.1\mu^2$) and broken 
($\lambda=50\mu^2$) phases, 
which are represented with diamonds and pluses, respectively. 
$J$ is increased at regular intervals. 
There is a one-to-one correspondence 
between the field configuration $\varphi(x^-)$ 
and the external field $J$. 
In the symmetric phase, 
the expectation value vanishes in the $J\to 0$ limit. 
In the broken phase, 
the curve tends to be closer to the $\varphi_{\rm max}$ axis 
with increasing harmonic resolution $K_{\rm cut}$. 
The classical field $\varphi(x^-)$ approaches the origin quickly 
as decreasing $J$. 
This fact suggests the presence of a nonzero 
field configuration $\varphi(x^-)$ 
of the $\lambda\phi_{1+1}^4$ model 
in the thermodynamic limit $L\to\infty$ ($K_{\rm cut}\to\infty$). 

In Fig. \ref{pot}, 
the effective energy ${\cal E}[\varphi]$ 
is plotted as a function of $\varphi_{\rm max}$ 
instead of as a functional of $\varphi(x^-)$. 
In the symmetric phase ($\lambda=0.1\mu^2$), 
we can see that the energy ${\cal E}[\varphi]$
has a minimum at the origin, 
where the state is composed 
only of a zero-momentum state $|0\rangle$ (Fock vacuum). 
The ground state of the hamiltonian $H[J]$ 
goes to $|0\rangle$ and gives zero energy 
${\cal E}[\varphi]=0$ in the $J\to 0$ limit. 
In the broken phase ($\lambda=50\mu^2$), however, 
a situation is completely different from the symmetric one. 
The effective energy has a flat bottom. 
A state on the flat region 
has a wavefunction where finite $K$ components are dominant 
because of a twist at the boundary. 
The classical field which is placed in the edge of the bottom 
jumps to zero in the $J\to 0$ limit. 
The classical field shows a singular behavior 
if the symmetry breaks spontaneously. 
This fact supports the existence of 
an infinite number of configurations $\varphi(x^-)$, 
which are energetically equivalent, 
and a nonzero field configuration 
as a kink solution of the Euler-Lagrange equation (\ref{el}) 
in the broken phase. 

\section{Summary and discussions}
\label{sec_sum}
We have found an approximate value 
of the critical coupling constant 
and obtained the effective energy by using DLCQ. 
In the symmetric phase, 
the effective energy has a minimum at the origin, 
which is composed only of the trivial Fock vacuum. 
In the broken phase, 
$Z_2$ symmetry spontaneously breaks, 
which has been confirmed by seeing that 
the bottom of the effective energy is flat. 
In the vanishing $J$ limit, 
a nonzero expectation value of the field seems to remain. 
A field configuration, which has a twist, 
can be a solution of the quantum mechanically extended 
Euler-Lagrange equation. 

In Sec. \ref{sec_dlcq}, 
mass spectrum calculation of particle states 
has been done with DLCQ. 
In the critical region, 
convergence of the spectra is very slow 
because the coupling constant is large there. 
This is due to an insufficiency of the harmonic resolution 
to represent small $k^+$ components of wavefunctions. 
By plotting the three-body wavefunction,  
we can see that 
the wavefunction increases rapidly at $k^+\sim 0$. 
A small finite $K$ value cannot represent such a sharp increase 
because the resolution is not sufficient. 
It would be hopeless to extend the DLCQ method 
by brute force to realistic models such as QCD, 
which has higher dimensions, 
since we couldn't get convergent result 
near the critical region 
even in this two-dimensional model. 
We need to renormalize the degrees of freedom 
of the harmonic resolution $K$ \cite{sande}. 
It would be better if we could also construct 
an effective hamiltonian 
by renormalizing higher Fock space truncated 
by the number of particles $N_{\rm TD}$ \cite{ho}. 

In the latter part of this paper, 
we have discussed spontaneous symmetry breaking 
by searching for a state that minimizes the effective energy. 
We have succeeded in finding an indication of 
spontaneous symmetry breaking, 
which is just contained in the hamiltonian 
with antiperiodic boundary condition. 
This suggests that the hamiltonian knows 
the existence of symmetry breaking 
in spite of an absence of a zero mode. 
We have considered how to extract information about 
symmetry breaking from the effective energy 
even though it is natural 
to use an effective potential for such investigations. 
It is easier to evaluate the effective energy 
than the effective potential 
because there is no vacuum fluctuations 
and the truncation of Fock space with respect to 
particle numbers works well in the light-front field theory. 

If the order parameter one would like to consider 
is a vacuum expectation value of 
a composite field such as 
$\langle 0| \bar{\psi}(x)\psi(x) |0 \rangle$, 
it is possible to trace spontaneous symmetry breaking 
by using a hamiltonian that has an interaction 
between the composite operator and an external field. 
There are a couple of possibilities to figure out 
symmetry breaking for the composite operator. 
One way is to define the effective potential 
with the zero mode of the composite operator 
$\bar{\psi}\psi$. 
Another way is to find a non-uniform (kink) solution 
of a classical field 
$\langle 0_J| \bar{\psi}\psi|0_J \rangle$ 
through the effective energy as discussed in this paper. 
To do that, we use a trick on $J(x)$, 
because the operator $\bar{\psi}\psi$ 
is always periodic even if any kinds of boundary conditions 
are imposed on the fermionic field $\psi$. 
Since the expectation value of the periodic operator 
can only have an even number of kinks, 
we have to assume an even number of kinks 
also on the external field.

\acknowledgments
It is my pleasure to thank K. Harada, K. Itakura, T. Matsuki, 
M. Taniguchi and M. Yahiro for helpful discussions 
and  all the members of Research Center for Nuclear Physics 
of Osaka university for their kind hospitality. 
I would also like to thank V. A. Miransky and K. Yamawaki 
for comments and I. S. Towner for correcting 
the whole manuscript. 
This work has been partially supported 
by the COE research fellowship. 

\appendix
\section{Partition function and Hamiltonian}
\label{p_h}
Let us prove that 
the following relation holds for the partition function $Z[J]$ 
when the external field is static $J(x)=J({\bdx})$ \cite{kugo}, 
\begin{equation}
  Z[J]
  \equiv
  \langle 0|
  Te^
  {i\int d^nx J(x)\phi(x)}
  |0 \rangle
  =\langle 0|e^{-iH[J]T}|0\rangle, 
  \label{a1}
\end{equation}
where 
\begin{equation}
  H[J] = H + H_J, 
  \quad H_J \equiv -\int d^{n-1}x J({\bdx})\bphi({\bdx}), 
\end{equation}
and ${\bdx}$ indicates $(n-1)$-dimensional spatial coordinates 
and $\bphi$ is a field operator represented 
in the Schr\"odinger picture. 
In order to prove the formula (\ref{a1}), 
the interaction picture will be defined 
regarding $H_J$ as a perturbed part. 

The Schr\"odinger equation for 
a general state vector is 
\begin{equation}
  i\frac{\partial}{\partial t}|\Psi_{\rm S}(t)\rangle 
  =
  H[J]|\Psi_{\rm S}(t)\rangle. 
  \label{sch}
\end{equation}
and a formal solution is readily obtained by writing 
\begin{equation}
  |\Psi_{\rm S}(t)\rangle
  =e^{-iH[J](t-t_0)}|\Psi_{\rm S}(t_0)\rangle. 
  \label{sch_kai}
\end{equation}
Define the interaction state vector in the following way 
\begin{equation}
  |\Psi_{\rm I}(t)\rangle = e^{iHt}|\Psi_{\rm S}(t)\rangle. 
  \label{con}
\end{equation}
The equation of motion of this state is easily found 
by carrying out the time derivative 
\begin{equation}
  i\frac{\partial}{\partial t}|\Psi_{\rm I}(t)\rangle 
  =
  H_{\rm I}(t)|\Psi_{\rm I}(t)\rangle, 
  \label{nise}
\end{equation}
\begin{equation}
  H_{\rm I}(t)
  \equiv e^{iHt} H_J e^{-iHt}
  =
  -\int d^{n-1}x J({\bdx})\phi(t,{\bdx}), 
\end{equation}
where $\phi(t,{\bdx})$ is the field operator 
in the interaction picture. 
The solution of Eq. (\ref{nise}) is 
\begin{equation}
  |\Psi_{\rm I}(t)\rangle
  =
  T \exp
  \left(
    i\int_{t_0}^t ds d^{n-1}x J({\bdx}) \phi(s,{\bdx})
  \right)
  |\Psi_{\rm I}(t_0)\rangle, 
  \label{nise_kai}
\end{equation}
where the time-ordered product is used 
due to the time dependence of the hamiltonian $H_{\rm I}(t)$. 
By substituting (\ref{sch_kai}) and (\ref{nise_kai})
into (\ref{con}), we have 
\begin{equation}
  e^{-iH[J](t_1-t_0)}
  =
  e^{-iHt_1}
  T\exp
  \left(
    i\int_{t_0}^{t_1} dt d^{n-1}x J({\bdx})\phi(t,{\bdx})
  \right), 
  \label{hoge}
\end{equation}
where 
$|\Psi_{\rm I}(t_0)\rangle = |\Psi_{\rm S}(t_0)\rangle$ 
is used. 
By sandwiching Eq. (\ref{hoge}) 
with the ground state $|0\rangle$ of the hamiltonian $H$, 
which satisfies $H|0\rangle = 0$, 
and setting $t_1=-t_0=T/2$ ($T$ is assumed to be large), 
we have 
\begin{equation}
  \langle 0|e^{-iH[J]T}|0 \rangle 
  =
  \langle 0|T\exp
  \left(
    i\int_{-T/2}^{T/2} dt d^{n-1}x J({\bdx})\phi(t,{\bdx})
  \right)
  |0\rangle. 
\end{equation}

\section{Hamiltonian}
\label{app_a}
The unperturbed ${\cal H}$ and 
the perturbed ${\cal H}_J$ parts of the hamiltonian (\ref{hj2}) 
\begin{equation}
  {\cal H}[J] = \frac{2\pi}{L} H[J]={\cal H} + {\cal H}_J, 
\end{equation}
is expressed using creation and annihilation operators 
in the two-dimensional $\lambda\phi^4$ model. 
The unperturbed part ${\cal H}$ is 
\begin{eqnarray}
  {\cal H} &=&
  \sum_{n=1}^\infty \frac{\mu^2}{\tilde{n}} a_n^\dagger a_n
  + \frac{1}{4}\frac{\lambda}{4\pi}
  \sum_{n_1,n_2,n_3,n_4=1}^\infty
  \frac{\delta_{\tilde{n}_1+\tilde{n}_2,\tilde{n}_3+\tilde{n}_4}}
  {\sqrt{\tilde{n}_1\tilde{n}_2\tilde{n}_3\tilde{n}_4}}
  a_{n_1}^\dagger a_{n_2}^\dagger a_{n_3} a_{n_4}
  \\
  && + \frac{1}{6}\frac{\lambda}{4\pi}
  \sum_{n_1,n_2,n_3,n_4=1}^\infty
  \frac{\delta_{\tilde{n}_1+\tilde{n}_2+\tilde{n}_3,\tilde{n}_4}}
  {\sqrt{\tilde{n}_1\tilde{n}_2\tilde{n}_3\tilde{n}_4}}
  [a_{n_1}^\dagger a_{n_2}^\dagger a_{n_3}^\dagger a_{n_4} + {\rm h.c.}]. 
\end{eqnarray}
and the harmonic resolution is 
\begin{equation}
  {\cal K} = \sum_{n=1}^\infty \tilde{n} a_n^\dagger a_n, 
\end{equation}
where
\begin{equation}
  \tilde{n} = n-\frac{1}{2}. 
\end{equation}

Even if the external field is assumed to be constant $J(x^-)=J$, 
$J(x^-)$ is not continuous at the boundary $x^-=\pm L$ 
because of the antiperiodicity of the field operator $\phi(x)$. 
This discontinuity is reflected as a twist 
in the expectation value of the field operator $\phi(x)$. 
It is possible to shift the position of the kink 
by taking the following external field, 
\begin{equation}
  J(x^-)=\cases{
    -J (-L\le x^- <a) \cr
    J (a<x^-<L) \cr}
\end{equation}
where $a$ indicates the position of the kink, $-L \le a < L$. 
The perturbed part of ${\cal H}[J]$ is given by 
\begin{equation}
  {\cal H}_J =
  -\sum_{n=1}^\infty \left( f_n(a) a_n^\dagger + f_n^*(a) a_n \right)
  \label{int_j}
\end{equation}
where
\begin{equation}
  f_n(a) \equiv \frac{2J}{\sqrt{\pi}\tilde{n}^{3/2}}
  \left[
    -  \sin \left( \frac{\pi a}{L}\tilde{n} \right)
    + i\cos \left( \frac{\pi a}{L}\tilde{n} \right)
  \right]. 
  \label{f_n}
\end{equation}
Of course, physics should be independent of the position $a$. 
It has been numerically confirmed 
that the ground state eigenvalue of $H[J]$ is independent of $a$.

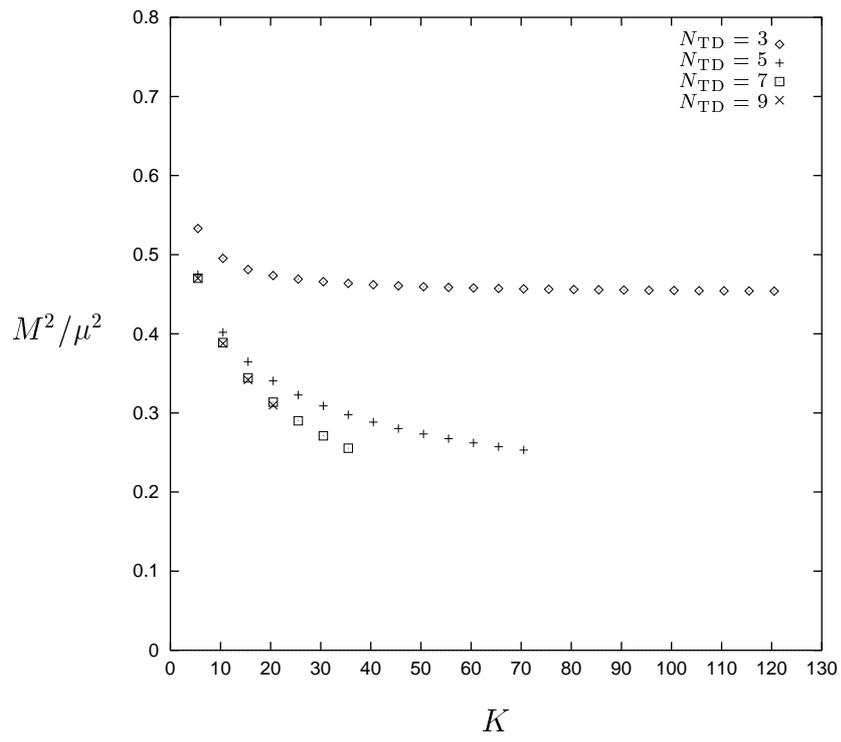
\begin{figure}[h]
  \caption[]{
    Mass squared $M^2$ of the lightest state is plotted 
    as a function of the harmonic resolution $K$ 
    for various Tamm-Dancoff truncations $N_{\rm TD}=3,5,7,9$
    under antiperiodic boundary conditions. 
    The coupling constant is taken as $\lambda=25\mu^2$, 
    which is relatively large 
    and near the critical point $\lambda_{\rm c}\sim 30\mu^2$. 
    The definite value of the critical coupling constant 
    $\lambda_{\rm c}$ will be calculated later. 
    The spectrum almost converges at $N_{\rm TD}=7$, 
    which can be confirmed only in the small $K$ region. 
    Mass spectrum calculations will be executed 
    in Fock space truncated with $K=141/2$ and $N_{\rm TD}=5$. 
    }
  \label{td}
\end{figure}
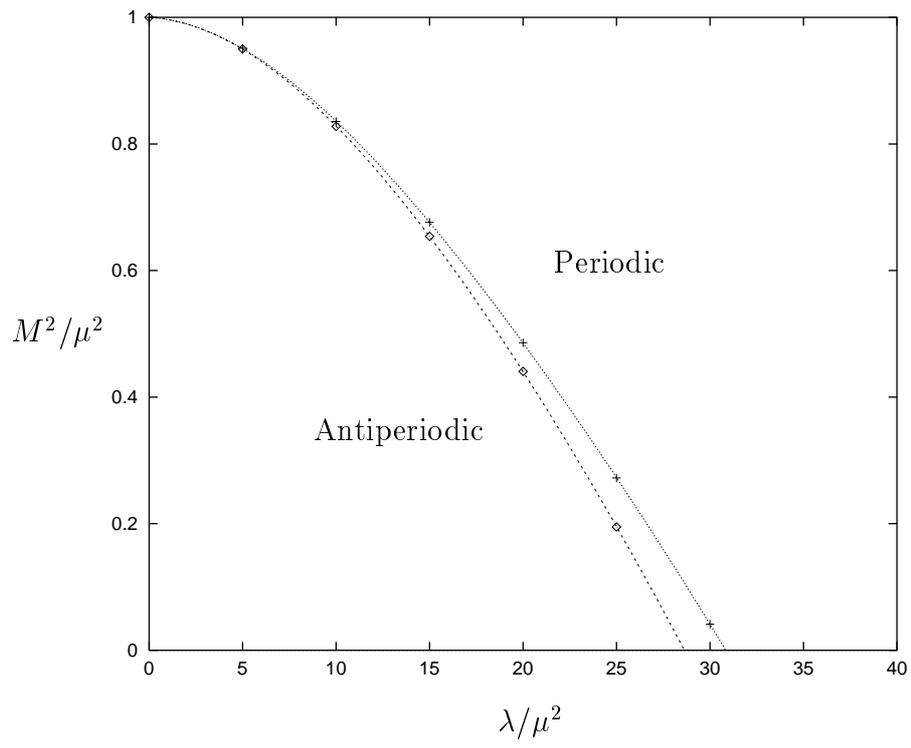
\begin{figure}[h]
  \caption[]{
    The extrapolated value $m_0$ of the lightest state is plotted 
    as a function of the coupling constant $\lambda$ 
    for antiperiodic and periodic boundary conditions, 
    each of which is represented 
    with diamonds and pluses, respectively. 
    The lines are intended to guide the eye and used 
    to calculated the critical coupling constant $\lambda_{\rm c}$. 
    In the periodic case, 
    the zero mode has been omitted 
    and the extrapolated values of the lightest state are 
    obtained from calculations in Fock space 
    truncated with $K=70$ and $N_{\rm TD}=5$ 
    in the same manner as the antiperiodic case. 
    The critical coupling constants are calculated as 
    $\lambda_{\rm c}=28.6329\mu^2$ (antiperiodic) and 
    $\lambda_{\rm c}=30.8431\mu^2$ (periodic)
    searching for the massless point $M^2(\lambda_{\rm c})=0$. 
    }
  \label{mass}
\end{figure}
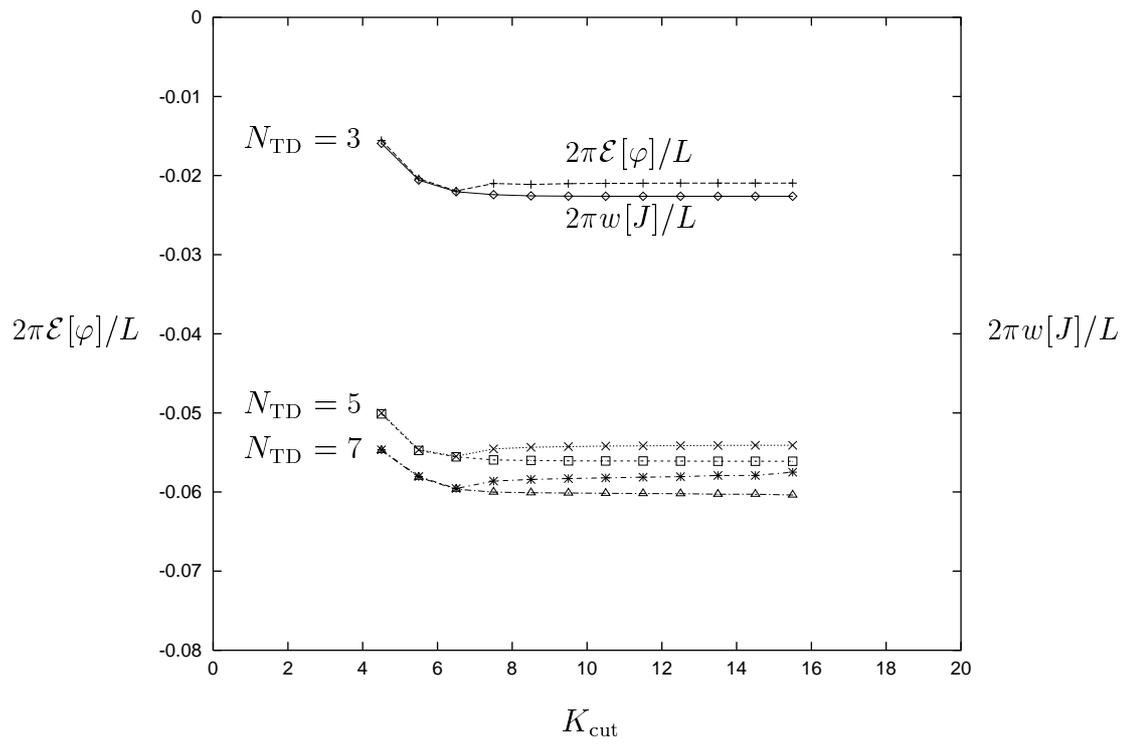
\begin{figure}[h]
  \caption[]{
    The effective energy ${\cal E}[\varphi]$ and 
    the generating functional $w[J]$ are plotted as functions 
    of $K_{\rm cut}$ for various Tamm-Dancoff truncations 
    $N_{\rm TD}=3,5,7$ in the broken phase, 
    where $\lambda=50\mu^2$ and $J=0.01$ are taken. 
    The effective energy ${\cal E}[\varphi]$ 
    always takes a larger value than $w[J]$. 
    A truncation with $K_{\rm cut}=31/2$ and $N_{\rm TD}=7$ 
    seems to give a spectrum that is near the convergent point. 
    This parameter set will be used in the subsequent calculations. 
    }
  \label{energy}
\end{figure}
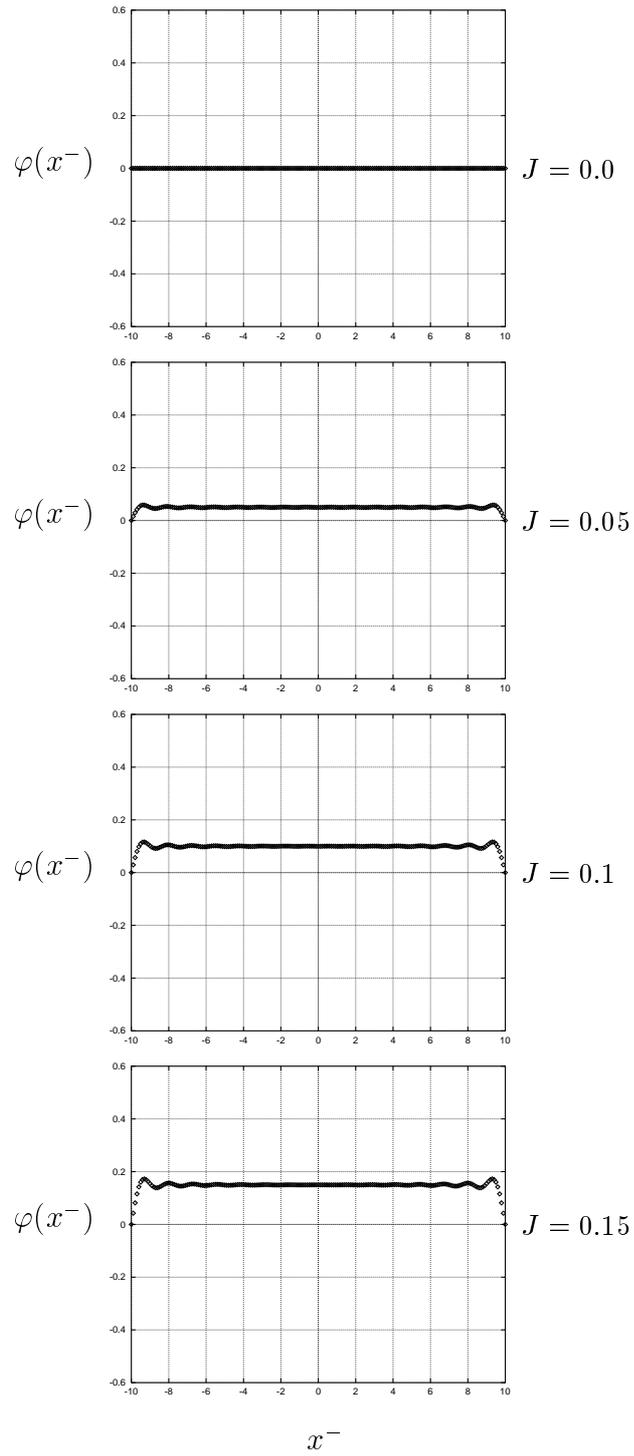
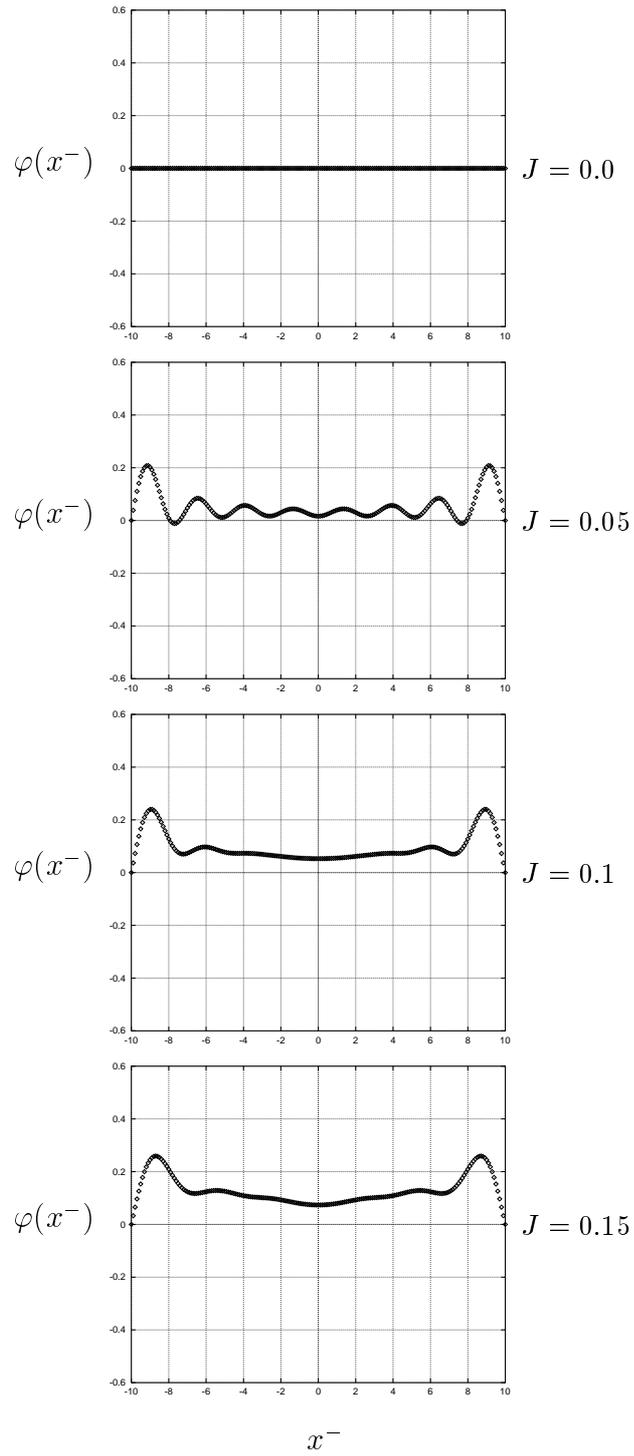
\begin{figure}[h]
  \caption[]{
    The classical field 
    $\varphi(x^-)=\langle 0_J|\phi(x^-)|0_J \rangle$ is plotted 
    as a function of $x^-$ both in 
    (a) symmetric ($\lambda=0.1\mu^2$) 
    and (b) broken ($\lambda=50\mu^2$) phases 
    for $J=0, 0.05, 0.1$, and $0.15$, where Fock space is 
    truncated with $K=31/2$ and $N_{\rm TD}=7$. 
    The vertical and horizontal axes stand for 
    $\varphi(x^-)$ and $x^-$ respectively. 
    We can observe that the classical field $\varphi(x^-)$ 
    has a twist at the boundary $x^-=\pm L$ 
    due to the antiperiodicity of the field operator $\phi(x)$. 
    }
  \label{phi_x}
\end{figure}
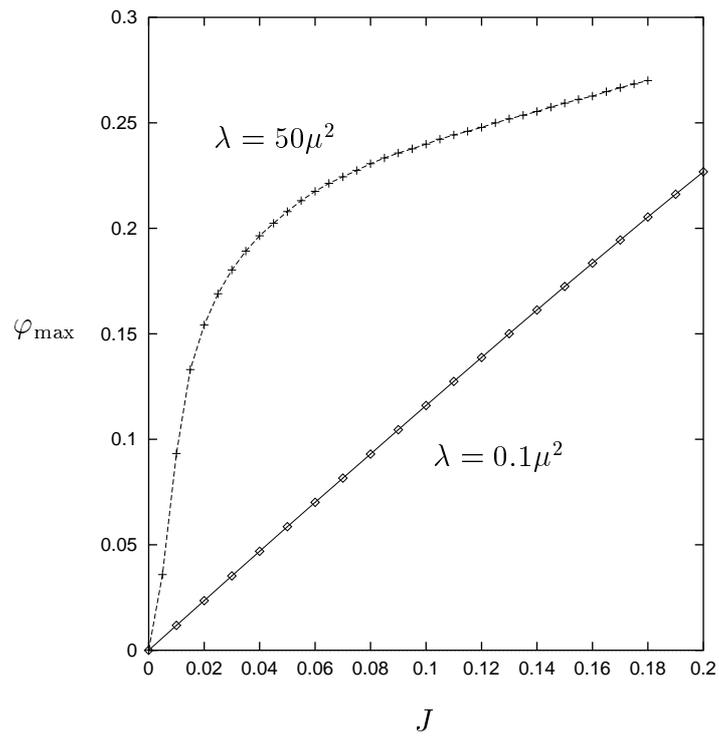
\begin{figure}[h]
  \caption[]{
    The maximum value 
    $\varphi_{\rm max}\equiv{\rm max}\{\varphi(x^-)\}$ 
    of the non-uniform classical field $\varphi(x^-)$ 
    is plotted as a function of $J$, 
    where Fock space is truncated 
    with $K_{\rm cut}=31/2$ and $N_{\rm TD}=7$. 
    Diamonds and pluses correspond to 
    symmetric ($\lambda=0.1\mu^2$) and 
    broken ($\lambda=50\mu^2$) phases, respectively. 
    The external field $J$ is changed 
    with an interval $\Delta J=0.005$. 
    In the broken phase, 
    the magnitude of the classical field $\varphi(x^-)$ 
    rapidly approaches the origin as $J$ decreases. 
    This suggests the presence of a nonzero field configuration 
    $\varphi(x^-)$ in the thermodynamic limit 
    $L\to\infty$ ($K_{\rm cut}\to\infty$). 
    }
  \label{jphi}
\end{figure}
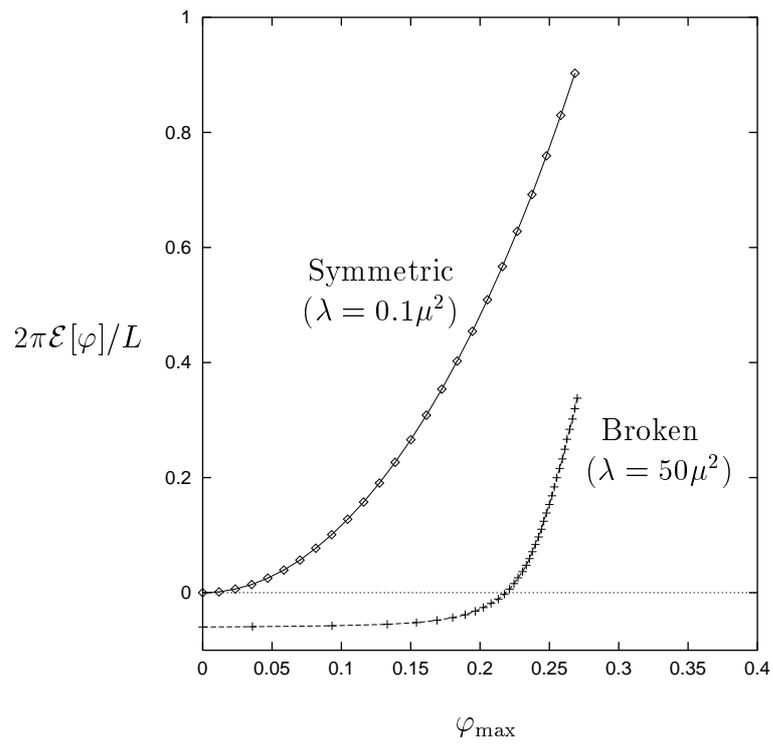
\begin{figure}[h]
  \caption[]{
    The effective energy $2\pi{\cal E}[\varphi]/L$ is plotted 
    as a function of $\varphi_{\rm max}$ 
    instead of as a functional of $\varphi(x^-)$, 
    where Fock space is truncated 
    with $K_{\rm cut}=31/2$ and $N_{\rm TD}=7$. 
    Diamonds and pluses correspond to 
    symmetric ($\lambda=0.1\mu^2$) and 
    broken ($\lambda=50\mu^2$) phases, respectively. 
    In the symmetric phase, 
    a physically meaningful configuration is at the origin, 
    where the state is composed only of the Fock vacuum $|0\rangle$. 
    In the broken phase, there seems to exist a nonzero 
    field configuration as a solution of the extended Euler-Lagrange 
    equation in the $J\to 0$ limit, 
    since the bottom of the effective energy is flat. 
    This says the existence of an infinite number of configurations 
    $\varphi(x^-)$ which are energetically equivalent. 
    }
  \label{pot}
\end{figure}

\begin{table}
\caption{
The $K$ dependence of the lightest one-bosonic mass $M^2$ 
is shown for various coupling constants, 
$\lambda/\mu^2=5, 10, 15, 20, 25$. 
The mass spectra are calculated in Fock space truncated 
with $N_{\rm TD}=5$ under antiperiodic boundary conditions. 
}
\begin{tabular}{r|ccccc}
  & & & $\lambda/\mu^2$ & \\
  $2K$ & 5 & 10 & 15 & 20 & 25 \\
  \tableline
  $11$ &
  $0.95800$ &
  $0.86614$ &
  $0.74880$ &
  $0.61655$ &
  $0.47481$ \\
  $21$ &
  $0.95471$ &
  $0.85293$ &
  $0.71987$ &
  $0.56736$ &
  $0.40190$ \\
  $31$ &
  $0.95341$ &
  $0.84712$ &
  $0.70629$ &
  $0.54317$ &
  $0.36476$ \\
  $41$ &
  $0.95268$ &
  $0.84368$ &
  $0.69787$ &
  $0.52769$ &
  $0.34045$ \\
  $51$ &
  $0.95221$ &
  $0.84134$ &
  $0.69196$ &
  $0.51657$ &
  $0.32267$ \\
  $61$ &
  $0.95188$ &
  $0.83962$ &
  $0.68750$ &
  $0.50804$ &
  $0.30883$ \\
  $71$ &
  $0.95163$ &
  $0.83829$ &
  $0.68398$ &
  $0.50120$ &
  $0.29763$ \\
  $81$ &
  $0.95144$ &
  $0.83722$ &
  $0.68110$ &
  $0.49555$ &
  $0.28829$ \\
  $91$ &
  $0.95129$ &
  $0.83634$ &
  $0.67870$ &
  $0.49079$ &
  $0.28034$ \\
  $101$ &
  $0.95116$ &
  $0.83560$ &
  $0.67666$ &
  $0.48669$ &
  $0.27346$ \\
  $111$ &
  $0.95105$ &
  $0.83497$ &
  $0.67489$ &
  $0.48312$ &
  $0.26744$ \\
  $121$ &
  $0.95096$ &
  $0.83442$ &
  $0.67334$ &
  $0.47997$ &
  $0.26209$ \\
  $131$ &
  $0.95088$ &
  $0.83394$ &
  $0.67196$ &
  $0.47716$ &
  $0.25731$ \\
  $141$ &
  $0.95081$ &
  $0.83351$ &
  $0.67074$ &
  $0.47465$ &
  $0.25301$ \\
\end{tabular}
\label{kdep_anti}
\end{table}

\begin{table}
\caption{
The $K$ dependence of the lightest one-bosonic mass $M^2$ 
is shown for various coupling constants, 
$\lambda/\mu^2=5, 10, 15, 20, 25$. 
The mass spectra are calculated in Fock space truncated 
with $N_{\rm TD}=5$ under periodic boundary conditions, 
where the zero mode has been removed from the hamiltonian. 
}
\begin{tabular}{r|ccccc}
  & & & $\lambda/\mu^2$ & \\
  $K$ & 5 & 10 & 15 & 20 & 25 \\
  \tableline
  $10$ &
  $0.97090$ &
  $0.90555$ &
  $0.82058$ &
  $0.72361$ &
  $0.61871$ \\
  $15$ &
  $0.96607$ &
  $0.88968$ &
  $0.78996$ &
  $0.67574$ &
  $0.55186$ \\
  $20$ &
  $0.96326$ &
  $0.88025$ &
  $0.77142$ &
  $0.64636$ &
  $0.51036$ \\
  $25$ &
  $0.96139$ &
  $0.87385$ &
  $0.75862$ &
  $0.62581$ &
  $0.48105$ \\
  $30$ &
  $0.96004$ &
  $0.86915$ &
  $0.74909$ &
  $0.61031$ &
  $0.45873$ \\
  $35$ &
  $0.95902$ &
  $0.86551$ &
  $0.74160$ &
  $0.59802$ &
  $0.44089$ \\
  $40$ &
  $0.95821$ &
  $0.86259$ &
  $0.73552$ &
  $0.58794$ &
  $0.42613$ \\
  $45$ &
  $0.95755$ &
  $0.86018$ &
  $0.73045$ &
  $0.57945$ &
  $0.41361$ \\
  $50$ &
  $0.95700$ &
  $0.85814$ &
  $0.72612$ &
  $0.57215$ &
  $0.40279$ \\
  $55$ &
  $0.95653$ &
  $0.85640$ &
  $0.72237$ &
  $0.56578$ &
  $0.39329$ \\
  $60$ &
  $0.95613$ &
  $0.85488$ &
  $0.71908$ &
  $0.56015$ &
  $0.38485$ \\
  $65$ &
  $0.95578$ &
  $0.85355$ &
  $0.71616$ &
  $0.55513$ &
  $0.37727$ \\
  $70$ &
  $0.95548$ &
  $0.85236$ &
  $0.71355$ &
  $0.55060$ &
  $0.37041$ \\
\end{tabular}
\label{kdep_peri}
\end{table}

\begin{table}
\caption{
The coefficients $m_i$ ($i=0,1,2$) 
in the $1/K$ expansion (3.17) 
are obtained from a least squares fitting 
with the Marquardt-Levenberg algorithm 
by using a result shown in TABLE I. 
$m_0$ is regarded as the extrapolated value of the lightest state 
in the thermodynamic limit $K\to\infty$. 
This is the result under antiperiodic boundary conditions. 
}
\begin{tabular}{r|ccc}
  $\lambda/\mu^2$ & $m_0$ & $m_1$ & $m_2$ \\
  \tableline
  $5$ &
  $ 0.95007    \pm 0.00002 $ &
  $ 0.05599    \pm 0.00082 $ &
  $-0.06878    \pm 0.00432 $ \\
  $10$ &
  $ 0.82974    \pm 0.00022 $ &
  $ 0.30798    \pm 0.00824 $ &
  $-0.59729    \pm 0.04307 $ \\
  $15$ &
  $ 0.66113    \pm 0.00077 $ &
  $ 0.82242    \pm 0.02820 $ &
  $-1.88593    \pm 0.14742 $ \\
  $20$ &
  $ 0.45652    \pm 0.00175 $ &
  $ 1.60482    \pm 0.06351 $ &
  $-4.01797    \pm 0.33191 $ \\
  $25$ &
  $ 0.22399    \pm 0.00317 $ &
  $ 2.63785    \pm 0.11523 $ &
  $-6.97948    \pm 0.60219 $ \\
\end{tabular}
\label{ml}
\end{table}


\begin{thebibliography}{30}
\bibitem{lapack}
  LAPACK, a library for numerical problems in linear algebra, 
  is used to perform diagonalization of matrices.
\end{thebibliography}
\end{document}